\documentclass[11pt]{article}
\usepackage[USenglish]{babel}
\usepackage{a4wide}
\usepackage{amsmath}
\usepackage{amssymb}
\usepackage{latexsym}
\usepackage{pifont}
\usepackage[all]{xy}
\usepackage{color}
\usepackage{pdfsync}
\usepackage{enumerate}
\usepackage{multicol}
\usepackage{wrapfig}
\usepackage[parfill]{parskip}
\usepackage{xr}
\usepackage{tikz}
\usepackage{verbatim}
\usepackage{float}
\usepackage{fancyhdr}
\usepackage[utf8]{inputenc}
\usepackage[T1]{fontenc}
\title{Congruences for Stochastic Automata}
\author{Ernst-Erich
  Doberkat\footnote{\texttt{doberkat@acm.org}}}
\date{\today}

\newcommand{\labelImpl}[2]{\ensuremath{\ref{#1}~\Rightarrow~\ref{#2}}}

\newcommand{\Klasse}[2]{\left[#1\right]_{#2}}
\newcommand{\Faktor}[2]{{#1}/{#2}}
\newcommand{\fMap}[1]{\eta_{#1}}
\newcommand{\Bild}[2]{{#1}\left[#2\right]}
\newcommand{\InvBild}[2]{\Bild{#1^{-1}}{#2}}
\newcommand{\Kern}[1]{\mathsf{ker}\left(#1\right)}

\newcommand{\SubProb}[1]{\spaceFont{S}\left(#1\right)}

\newcommand{\SubProbSenza}{\spaceFont{S}}

\newcommand{\Borel}[1]{\ensuremath{{\mathcal B}(#1)}}

\newtheorem{definition}{Definition}[section]
\newcommand{\BeginDefinition}[1]{%
  \begin{definition}\label{#1}
}
\newcommand{\EndDefinition}{\end{definition}}

\newtheorem{example}[definition]{Example}
\newcommand{\BeginExample}[1]{%
  \begin{example}\label{#1}\rm
}
\newcommand{\EndExample}{--- \end{example}}

\newtheorem{observation}[definition]{Observation}
\newcommand{\BeginObservation}[1]{
  \begin{observation}\label{#1}\rm
}
\newcommand{\EndObservation}{--- \end{observation}}

\newtheorem{theorem}[definition]{Theorem}
\newcommand{\BeginTheorem}[1]{%
  \begin{theorem}\label{#1}
}
\newcommand{\EndTheorem}{\end{theorem}}

\newtheorem{corollary}[definition]{Corollary}
\newcommand{\BeginCorollary}[1]{
  \begin{corollary}\label{#1}
}

\newtheorem{proposition}[definition]{Proposition}
\newcommand{\BeginProposition}[1]{%
  \begin{proposition}\label{#1}
}
\newcommand{\EndProposition}{\end{proposition}}

\newcommand{\EndCorollary}{\end{corollary}}
\newtheorem{lemma}[definition]{Lemma}
\newcommand{\BeginLemma}[1]{%
  \begin{lemma}\label{#1}
}
\newcommand{\EndLemma}{\end{lemma}}

\newtheorem{claim}{Claim}
\newcommand{\BeginClaim}[1]{%
  \begin{claim}\label{#1}
}
\newcommand{\EndClaim}{\end{claim}}

\newenvironment{proof}{\textbf{Proof\ }}{\ensuremath{\QED}}
\newcommand{\BeginProof}{\begin{proof}}
\newcommand{\EndProof}{\end{proof}}

\newenvironment{remark}{\textbf{Remark:\ }}{}
\newcommand{\BeginRemark}{\begin{remark}}
\newcommand{\EndRemark}{\QED\end{remark}}
\newcommand{\QED}{%
\ensuremath{\dashv}
}

\newcommand{\Real}{\mathbb{R}}

\newcommand{\Nat}{\mathbb{N}}

\def\isEquiv#1#2#3{#1~#3~#2}
\def\GlueOpp{\ast}
\def\comb#1#2{#1\GlueOpp#2}
\def\chi{\xi}
\def\Dt#1{{#1}_{\bullet}}
\makeatletter
\def\@auf#1{{#1}^{\sharp}}
\def\@ab#1{{#1}^{\flat}}
\def\hin#1{\@auf{#1}}
\def\her#1{\@ab{#1}}
\makeatother
\def\rnd#1{{#1}^{\diamond}}
\def\eins#1{1_{#1}}
\renewcommand{\SubProbSenza}{\mathcal{G}}
\renewcommand{\SubProb}[1]{\SubProbSenza\left(#1\right)}
\renewcommand{\Kern}[1]{\mathfrak{ker}\left(#1\right)}
\renewcommand{\Borel}[1]{\ensuremath{{\mathfrak{B}}(#1)}}
\begin{document}
\maketitle
\begin{abstract}
  Congruences for stochastic automata are defined, the corresponding
  factor automata are constructed and investigated for automata over
  analytic spaces. We study the behavior under finite and infinite
  streams.
  Congruences consist
  of multiple parts, it is shown that factoring can be done in
  multiple steps, guided by these parts. \\[0.1cm]

\end{abstract}
\section{Introduction}
\label{sec:introduction}

Stochastic automata~\cite{Asendorpf,EED-LNCS81} are the natural generalization to
non-deterministic Mealy automata; they
take an input while being in an internal state, change their state and
return an output. Both the new state and the output are distributed
according to the automaton's
transition law. The basic
scenario may be finite or infinite, in the infinite case one may deal
with countable or uncountable carrier sets for input, outputs, and
states, resp. The finite and the countably infinite case is usually
delt with through methods from linear algebra, since matrices with a
finite or countable number of entries are manipulated, the uncountable case required
methods from measure theory. This is so since the events an automata
is assumed to handle are not all possible events, but come from Boolean $\sigma$-algebras of events
(using all possible events, i.e., defining the probabilities on the
respective power sets, will lead to foundational problems).

This kind of automata ---~without the bells and whistles one finds in
later extensions~--- have been used, e.g., for modelling simple
learning processes along a behavioral taxonomy from
psychology~\cite{Skinner, Klix, EED-LNCS81}. In such a scenario, in which the automaton models
a learner, the automaton receives inputs from the environment while
being in a specific state, it makes a state transition and responds
with an output. This happens in a sequential fashion. We are
interested in the single-step behavior. The learning situation is
characterized by the observation that equivalent inputs may lead to
equivalent outputs, and that there may be equivalent states as well;
note that the set of states represents an abstraction obtained through
a modelling process, hence is not accessible from the outside. For
conceptual clarity, and for minimizing the machine at least
conceptually, one is interested in these equivalences, i.e., one wants
to form equivalence classes and have the transition law respect these
classes. This leads to the notion of a \emph{congruence}, well known in
(universal) algebra. But we must not ignore a slightly inconvenient
fact: while a congruence, say, on a group, relates group elements
to each other, an automaton congruence relates pairs of inputs and
states to pairs of states and outputs, so we have a slightly
heterogeneous situation at hand. One might be reminded of
bisimilarity, where sets of two possibly different transition systems
are related to each other.

The latter problem is resolved by introducing the notion of \emph{friendship} for
two equivalence relations, comparing their probabilistic behavior 
in a straightforward manner. This leads to a notion of congruence for
automata, which is exploited by relating it to morphisms and their
kernels and constructing factor automata.

An automaton works sequentially, so we study the automaton's behavior
for finite and for infinite input sequences. Here we adopt a black box
point of view, hiding state changes from the outside world. This is
studied first for finite sequences, then we construct a limit which
permits us also to specify behavior under an infinite input stream. It
turns out that friendship is a surprisingly stable relationship which
can be maintained also for infinte streams.

Finally we want to know
whether we can form longer chains of reduced automata, and it turns
out that this is not possible: factoring a factored automaton yields
an automaton which can be obtained through one-step factoring through
a suitably modifies congruence. The result also enables us to reduce
automata in a step wise fashion along its components. 

Most of the material depends heavily on the coalgebraic approach to
stochastic relations~\cite{Panangaden, EED-Companion,
  EED-CoalgLogic-Book}. The present paper rests on the well-known fact that the
very old problem of reducing an automaton may be solved in a more
general fashion without much effort with tools from coalgebras~\cite{EED-Companion, Jacobs-coalg}.

\subsubsection*{Notation and all that}
\label{sec:notation-all-that}

A \emph{measurable space} $(F, \mathcal{F})$ is a set $F$ together with a Boolean
$\sigma$-algebra $\mathcal{F}$ of subsets of $F$. Measurable spaces form a category,
taking measurable maps as morphism. A map $f: F \to H$ for the
measurable spaces $(F, \mathcal{F})$ and $(H, \mathcal{H})$ is said to
be \emph{$\mathcal{F}$-$\mathcal{H}$ measurable} iff
$\InvBild{f}{\mathcal{H}} \subseteq \mathcal{F}$, i.e., iff $\InvBild{f}{Q}\in
\mathcal{F}$ holds for every $Q\in \mathcal{H}$; we will omit
the $\sigma$-algebras from the notation of maps whenever possible. The \emph{Giry functor} $\SubProbSenza$
acts as an endofunctor on this category. It assigns to each measurable
space $(F, \mathcal{F})$ the set $\SubProb{F, \mathcal{F}}$ of all
subprobabilities on $\mathcal{F}$ equipped with the smallest
$\sigma$-algebra rendering the evaluations $\mu\mapsto \mu(Q)$ for all
$Q\in \mathcal{F}$ measurable. To complete the definition of the functor
$\mathcal{G}$, map the measurable map $f: F\to H$ to the measurable
map $\SubProb{f}$ which assigns each subprobability $\mu$ on
$\mathcal{F}$ its image $\lambda P.\mu(\InvBild{f}{P})$ on
$\mathcal{H}$.

Assume an equivalence relation $\chi$ on the measurable space $(F,
\mathcal{F})$. The map $\eta_{\chi}: x\mapsto \Klasse{x}{\chi}$ sends an
element to its $\chi$-class. Denote as
usual the set of $\chi$-classes by $\Faktor{F}{\chi}$. This set will be furnished with the
$\sigma$-algebra $\Faktor{\mathcal{F}}{\chi}$ which is the final
$\sigma$-algebra on $\Faktor{F}{\chi}$ with respect to $\mathcal{F}$ and $\eta_{\chi}$,
thus $V\in \Faktor{\mathcal{F}}{\chi}$ iff
$\InvBild{\eta_{\chi}}{V}\in\mathcal{F}$. We denote the measurable space
$\bigl(\Faktor{F}{\chi}, \Faktor{\mathcal{F}}{\chi}\bigr)$ by
$\Faktor{(F, \mathcal{F})}{\chi}$. $\eins{F}$ denotes the identity
relation on $F$. 

\section{Stochastic Automata}
\label{stoch-aut}
A \emph{stochastic relation} $K:(X, \mathcal{A})\Rightarrow (Y, \mathcal{B})$ is a
measurable map $K: X\to\SubProb{Y, \mathcal{B}}$, thus $K(x)$ is a
subprobability measure on $(Y, \mathcal{B})$ for each $x\in X$, and
the map $x\mapsto K(x)(B)$ is $\mathcal{A}$-measurable for each
$B\in\mathcal{B}$. Actually ---~but inconsequentially for the present
note~--- a stochastic relation is a Kleisli
morphism for the Giry monad, the functorial part of which is the Giry
functor $\SubProbSenza$~\cite{EED-Companion, Panangaden-book}. 

Recall that the category of measurable spaces is
closed under finite products: $(X, \mathcal{A})\otimes(Y,
\mathcal{B})$ has the Cartesian product $X\times Y$ as a carrier set
and $\sigma(\{A\times B \mid A\in\mathcal{A}, B\in\mathcal{B}\}) =:
\mathcal{A}\otimes\mathcal{B}$ as a $\sigma$-algebra. Here
$\sigma(\{\dots\})$ denotes the smallest $\sigma$-algebra on the
carrier containing the generator $\{\dots\}$. A $\sigma$-algebra is
\emph{countably generated} iff it has a countable generator, and it \emph{separates
points} iff given two distinct points there is a measurable set
containing exactly one of them. It is well known that countably
generated, point separating $\sigma$-algebras are precisely the Borel
sets for second countable metric spaces~\cite{Srivastava}.

\BeginDefinition{Automaton}
A \emph{stochastic automaton} $\mathbf{K} = \bigl( (X, \mathcal{A}), (Y, \mathcal{B}), (Z,
\mathcal{C}), K \bigr)$ is a stochastic relation $K: (X\times Z,
\mathcal{A}\otimes\mathcal{C}) \Rightarrow (Z\times Y,
\mathcal{C}\otimes \mathcal{B})$.  
\EndDefinition
Thus the new state and the output of $\mathbf{K}$ is a
member of the measurable set $D\in\mathcal{C}\otimes\mathcal{B}$ with
probability $K(x, z)(D)$ upon input $x\in X$ in state $z\in
Z$. Because we work in the realm of subprobabilities, mass may get
lost, so that we cannot always reckon with $K(x, z)(Z\times Y) =
1$. This suggests the possibility that events cannot be accounted
for. 

The automata may work in different environments, so different input
and output spaces have to be taken into account. Morphisms are used
for relating automata. Assume that we have another stochastic
automaton $\mathbf{K}' = \bigl( (X', \mathcal{A}'), (Y', \mathcal{B}'), (Z',
\mathcal{C}'), K' \bigr)$. A morphism
$\mathfrak{f}: \mathbf{K}\to \mathbf{K'}$ is a triplet $\mathfrak{f} = (f, g, h)$
of surjective measurable
map $f: Z\to Z'$, $g: Y\to Y'$ and $h: Z\to Z'$ rendering this diagram
commutative (with, e.g.,  $f\times
h: \langle x, z\rangle\mapsto\langle f(x), h(z)\rangle$):
\begin{equation*}
\xymatrix{
X\times Z\ar[rr]^{K}\ar[d]_{f\times h}&&\SubProb{(Z\times Y, \mathcal{C}\otimes
  \mathcal{B}})\ar[d]^{\SubProb{h\times g}}\\
X'\times Z'\ar[rr]_{K'}&&\SubProb{(Z'\times Y', \mathcal{C'}\otimes
  \mathcal{B'}})
}
\end{equation*}
Thus
\begin{multline*}
K'(f(x), h(z))(E) =
  \bigl(K'\circ (f\times h)\bigr)(x, z)(E) =\\
\bigl(\SubProb{h\times g}\circ K\bigr)(x, z)(E) =
  K(x, z)(\InvBild{(h\times g)}{E})
\end{multline*}
whenever $E\in\mathcal{C}'\otimes\mathcal{B}'$ indicates the operation of automaton $\mathbf{K}'$.

\section{Congruences}
\label{congruences}

Before we define congruences for stochastic automata, we need to talk
about friendly relations, i.e., relations on different states which
behave nevertheless like congruences. To be specific:
Given a stochastic relation $K: (F, \mathcal{F})\Rightarrow (H,
\mathcal{H})$ and equivalence relations $\chi$ and $\vartheta$ on $F$
resp. $H$, call \emph{$\chi$ friendly to $\vartheta$} iff there exists a
stochastic relation $K_{\chi, \vartheta}: \Faktor{(F,
  \mathcal{F})}{\chi}\Rightarrow \Faktor{(H, \mathcal{H})}{\vartheta}$
rendering this diagram commutative:
\begin{equation}
  \label{friend}
\xymatrix{
F\ar[rr]^{K}\ar[d]_{\eta_{\chi}}&&\SubProb{H, \mathcal{H}}\ar[d]^{\SubProb{\eta_{\vartheta}}}\\
\Faktor{F}{\chi}\ar[rr]_{K_{\chi, \vartheta}}&&\SubProb{\Faktor{(H, \mathcal{H})}{\vartheta}}
}
\end{equation}
We observe for friendly $\chi, \vartheta$ that
\begin{equation*}
K_{\chi, \vartheta}(\Klasse{x}{\chi})(T) =
\bigl(\SubProb{\eta_{\vartheta}}\circ K\bigr)(x)(T) = 
K(x)(\InvBild{\eta_{\vartheta}}{T},)
\end{equation*}
so that $\chi$ and $\vartheta$ indeed cooperate in
a congruential manner.

We will also need the concept of a small equivalence relation, given
that \emph{equivalence} is a very broad notion. It needs to be restricted
somewhat for being useful in our context.

Again, assume an equivalence relation $\chi$ on the measurable space $(F,
\mathcal{F})$. Call the set $Q\in \mathcal{F}$ \emph{$\chi$-invariant} iff $Q$ is the union
of equivalence classes, thus iff $x\in Q$ and
$\isEquiv{x}{x'}{\chi}$ entails $x'\in Q$. It is not difficult to
see that
\def\invar#1#2{[\mathcal{#1}, #2]}
\begin{equation}
  \label{INV}
  \invar{F}\chi := \{Q\in \mathcal{F} \mid Q\text{ is $\chi$-invariant}\}
\end{equation}
is a $\sigma$-algebra, the \emph{$\sigma$-algebra of
  $\chi$-invariant sets}. Observe that
$\Bild{\eta_{\chi}}{U}\in\Faktor{\mathcal{F}}{\chi}$ for
$U\in\invar{F}{\chi}$, because
$\InvBild{\eta_{\chi}}{\Bild{\eta_{\chi}}{U}} = U$. Call the
equivalence relation $\chi$
\emph{small} iff there exists a countable family
$\bigl(U_{n}\bigr)_{n\in \Nat}\subseteq\mathcal{F}$ such that
\begin{equation*}
  \isEquiv{x}{x'}{\chi}\text{ iff }\forall n\in \Nat: x\in
  U_{n}\Leftrightarrow x'\in U_{n}.
\end{equation*}
$\bigl(U_{n}\bigr)_{n\in\Nat}$ is said to \emph{create} relation
$\chi$. Then $\invar{\mathcal{F}}{\chi} = \sigma(\{U_{n} \mid n\in\Nat\})$ is
  countably generated, so is $\Faktor{\mathcal{F}}{\chi}$, which also separates
  points.

\BeginExample{kerf-smooth}
  Let $f: (F, \mathcal{F})\to(H, \mathcal{H})$ be measurable, and
  assume that
  $\mathcal{H}$ is countably generated and separates points. Then the
  kernel relation
  \begin{equation*}
      \Kern{f} := \{\langle x, x'\rangle \mid f(x) = f(x')\}
    \end{equation*}
  is small. In fact, let $(U_{n})_{n\in\Nat}$ be the generator for
  $\mathcal{H}$, then we show that $\{U_{n} \mid n\in\Nat\}$ separates
  points. Take $y, y'\in H$ such that $y\in
  U_{n}$ iff $y'\in U_{n}$ for all $n\in\Nat$. Since $\{U\subseteq H
  \mid \forall n\in\Nat: y\in U \Leftrightarrow y'\in U\}$ is a
  $\sigma$-algebra which contains the generator, it contains
  $\mathcal{H}$. From this we conclude that $y = y'$. But this means
  that $\bigl(\InvBild{f}{U_{n}}\bigr)_{n\in\Nat}$ creates $\Kern{f}$. 
\EndExample

The following observation helps characterizing friendly equivalence
relations.

\BeginLemma{charact}
Let $K: (F, \mathcal{F})\Rightarrow (H,
\mathcal{H})$ be a stochastic relation  and assume equivalence
relations $\chi$ and $\vartheta$ on $F$ resp. $H$, are given. Then these
conditions are equivalent:
\begin{enumerate}
\item\label{one} $\chi$ is friendly to $\vartheta$.
\item\label{two} $\SubProb{m_\vartheta}\circ K: (F, [\mathcal{F},
  \chi])\Rightarrow (H, [\mathcal{H}, \vartheta])$ with $m_{\vartheta}: (H,
  \mathcal{H})\to(H, [\mathcal{H}, \vartheta])$ as the identity.
\item\label{three} $\Kern{\SubProb{m_\vartheta}\circ K} \supseteq \chi$. 
\end{enumerate}
\EndLemma

\BeginProof
Abbreviate the map $\SubProb{m_\vartheta}\circ K$ by $L$, and note that
$\SubProb{m_\vartheta}$ restricts measures on $\mathcal{H}$ to its
sub $\sigma$-algebra $[\mathcal{H}, \vartheta]$.

\labelImpl{one}{two}: It is clear that $L: (F, \mathcal{F})\Rightarrow
(H, [\mathcal{H}, \vartheta])$, because $\SubProb{m_\vartheta}$ acts as
restriction to $[\mathcal{H}, \vartheta]$. So it has to be shown that $x\mapsto L(x)(G)$ is
$[\mathcal{F}, \chi]$-measurable for each $G\in[\mathcal{H},
\vartheta]$. Let $G_{0} :=
\Bild{\eta_{\vartheta}}{G}\in\Faktor{\mathcal{H}}{\vartheta}$, then
$L(x, G) = L(x, \InvBild{\eta_{\vartheta}}{G_{0}}) =
(\SubProb{\eta_{\vartheta}}\circ K)(x)(G_{0})$,
thus $L(x)(G) < r \text{ iff } K_{\chi,
 \vartheta}(\Klasse{x}{\chi})(G_{0}) < r$, which implies measurability
of $x\mapsto L(x)(G)$.

\labelImpl{two}{three}: The assumption that there exists
$T\in[\mathcal{H}, \vartheta]$ such that $K(x)(T) < r < K(x')(T)$ for some
$x, x'$ with $\isEquiv{x}{x}{\chi}$ gives immediately a
contradiction.

\labelImpl{three}{one}: Define
$ K_{\chi, \vartheta}(\Klasse{x}{\chi}) :=
(\SubProb{\eta_{\vartheta}}\circ K)(x)$, then $K_{\chi, \vartheta}$ is
well-defined, satisfies the measurability conditions and renders 
diagram (\ref{friend}) commutative. 
\EndProof

This useful characterization permits testing friendship without actually
constructing the factors. It extends to bounded, measurable functions:

\BeginCorollary{inv-under-integral}
Under the assumptions of Lemma~\ref{charact}, these statements are
equivalent
\begin{enumerate}
\item\label{one-1} $\chi$ is friendly to $\vartheta$.
\item\label{four-1} For each bounded and $[\mathcal{H},\vartheta]$-measurable $f: H\to
  \Real$
  \begin{equation*}
    \isEquiv{x}{x'}{\chi} \Rightarrow \int_{H}f~dK(x) = \int_{H}f~dK(x').
  \end{equation*}
\end{enumerate}
\EndCorollary

\BeginProof
The implication \labelImpl{one-1}{four-1} follows from
part~\ref{three} in Lemma~\ref{charact} together with the observation that
a bounded measurable function is the pointwise limit of a sequence of step
functions, and Lebesgue's Convergence Theorem. The converse
implication observes that the indicator function of a measurable set
is a bounded measurable function. An application of
part~\ref{three} in Lemma~\ref{charact} yields the result. 
\EndProof

An interesting example for friendship is given by kernels of morphisms for
stochastic relations. Recall that finality of a measurable map $f: (F,
\mathcal{F})\to(H,\mathcal{H})$ may be characterized by the property that
$\mathcal{H} = \{R\subseteq H \mid \InvBild{f}{R}\in\mathcal{F}\}$. Thus
we may conclude from
$\InvBild{f}{R}\in\mathcal{F}$ that $R\in \mathcal{H}$, provided $f$
is final and onto.

\BeginExample{morpf}
Let $K_{i}: (F_{i}, \mathcal{F}_{i})\Rightarrow
(H_{i},\mathcal{H}_{i})$ be stochastic relations for $i = 1, 2$, and
assume that $(f, g): K_{1}\to K_{2}$ is a morphism, which means 
$K_{2}\circ f = \SubProb{g}\circ K_{1}$ for the surjective measurable
maps $f: F_{1}\to F_{2}$ and $g: H_{1}\to H_{2}$. We claim that $\Kern{f}$ is friendly to $\Kern{g}$, provided
$g$ is final and onto.

In fact,
let $f(x) = f(x')$, then we have to show that $K_{1}(x)(G) =
K_{1}(x')(G)$ for all $G\in[\mathcal{H}_{1}, \Kern{g}]$. Fix such a
set $G$, then we know that 
$G = \InvBild{\eta_{\Kern{g}}}{\Bild{\eta_{\Kern{g}}}{G}}$
with
$\Bild{\eta_{\Kern{g}}}{G}\in\Faktor{\mathcal{H}_{1}}{\Kern{g}}$. Factoring $g = \Dt{g}\circ \eta_{\Kern{g}}$ with $\Dt{g}:
    \Faktor{H_{1}}{\Kern{g}}\to H_{2}$ measurable, final and injective yields
    the surjective map $\Dt{g}^{-1}$ between powersets. We find
      therefore $H_{0}\subseteq H_{2}$ with $\InvBild{\Dt{g}}{H_{0}} =
      \Bild{\eta_{\Kern{g}}}{G}$. Because 
\begin{equation*}
\InvBild{g}{H_{0}}
=
\InvBild{\eta_{\Kern{g}}}{\InvBild{\Dt{g}}{H_{0}}}
=
\InvBild{\eta_{\Kern{g}}}{\Bild{\eta_{\Kern{g}}}{G}}
=
G \in[\mathcal{H}_{1}, \Kern{g}]\subseteq \mathcal{H}_{1}
\end{equation*}
we conclude from finality of $\Dt{g}$ that $H_{0}\in\mathcal{H}_{2}$, so that
\begin{multline*}
  K_{1}(x)(G) =\\ K_{1}(x)(\InvBild{g}{H_{0}})
  =
  (\SubProb{g}\circ K_{1})(x)(H_{0})
    =
    K_{2}(f(x))(H_{0})
    =
    K_{2}(f(x'))(H_{0})
    =\\
    K_{1}(x')(G).
\end{multline*}
This gives the assertion.
\EndExample

After all these preparations we are in a position to define
congruences for stochastic automata.

\BeginDefinition{congr-ext}
Let $\mathbf{K} = \bigl( (X, \mathcal{A}), (Y, \mathcal{B}), (Z,
\mathcal{C}), K \bigr)$ be a stochastic automaton, then a triplet $\mathfrak{c} =
(\alpha, \beta, \gamma)$ of equivalence relations on $X, Y$ resp. $Z$
is called a\emph{ congruence for $\mathbf{K}$} iff $\alpha\times\gamma$ is friendly to
$\gamma\times\beta$. 
\EndDefinition

A congruence $\mathfrak{c}$ for stochastic automaton $\mathbf{K}$ is characterized by the
existence of a stochastic relation
\begin{equation}
  \label{congr-ext-1}
K_{\mathfrak{c}}: \Faktor{\bigl((X, \mathcal{A})\otimes(Z,
  \mathcal{C})\bigr)}{(\alpha\times\gamma)}
\Rightarrow
\Faktor{\bigl((Z, \mathcal{C})\otimes(Y,
  \mathcal{B})\bigr)}{(\gamma\times\beta)}
\end{equation}
which renders this diagram commutative:
\begin{equation*}
\xymatrix{
  (X, \mathcal{A})\otimes(Z,\mathcal{C})
  \ar[d]_{\eta_{\alpha\times\gamma}}
  \ar[rr]^{K}
&&
\SubProb{(Z, \mathcal{C})\otimes(Y,\mathcal{B})}
\ar[d]^{\SubProb{\eta_{\gamma\times\beta}}}\\
\Faktor{\bigl((X, \mathcal{A})\otimes(Z,
  \mathcal{C})\bigr)}{(\alpha\times\gamma)}
\ar[rr]_{K_{\mathfrak{c}}}
&&
\SubProb{\Faktor{\bigl((Z, \mathcal{C})\otimes(Y,
\mathcal{B})\bigr)}{(\gamma\times\beta)}}
}
\end{equation*}

This is an immediate consequence:

\BeginProposition{factor-is-morph}
In the notation of Definition~\ref{congr-ext}, $(\eta_{\alpha},
\eta_{\beta}, \eta_{\gamma}): \mathbf{K}\to \mathbf{K}_{\mathfrak{c}}$ is a morphism. \QED
\EndProposition

The classic case of state reduction by a relation $\gamma$ for
automaton $\mathbf{K} = \bigl( (X, \mathcal{A}), (Y, \mathcal{B}), (Z,
\mathcal{C}), K \bigr)$ is captured
through the triplet $\mathfrak{s} = (\eins{X}, \eins{Y}, \gamma)$;
that $\mathfrak{s}$ is a congruence for $\mathbf{K}$ is characterized through
\begin{equation*}
\forall B\in{\cal B}: K(x, z)(E\times B) = K(x, z')(E\times B),
\end{equation*}
whenever $\isEquiv{z}{z'}{\gamma}$, and $E\in[{\cal C}, \gamma]$ is a
$\gamma$-invariant measurable subset of $Z$. This is quite close to
the intuition of a (state-) congruence for an automaton: equivalent
states behave in the same way on measurable sets which cannot separate
equivalent states.

On the other hand, one probably wants to leave the states alone and
cater only for inputs and outputs. Here one would work with $\mathfrak{t} =
(\alpha, \beta, \eins{Z})$, and $\mathfrak{t}$ is a congruence iff
\begin{equation*}
  \forall C\in{\cal C}: K(x, z)(C\times B) = K(x', z)(C\times B),
\end{equation*}
whenever $\isEquiv{x}{x'}{\alpha}$ and $B\in[{\cal B}, \beta]$, so the
behavior of $\mathbf{K}$ on inputs which are identified through
$\alpha$ is the same on sets which cannot separate $\beta$-equivalent
outputs. Certainly other combinations are possible.  

It is noted that the behavior of an automaton is completely
characterized by its assigning values to sets of the form $C\times
B$. This is so because these sets determine the respective
product $\sigma$-algebras uniquely, and their collection is closed
under intersections~\cite[Lemma 1.6.31]{EED-Companion}.

\section{Factoring}
\label{sec:factoring}

We will restrict the class of measurable spaces to analytic spaces now,
and we will deal only with small equivalence relations.

Recall that an
\emph{analytic space} is the measurable image of a \emph{Polish space}, i.e., of a
second countable, completely metrizable topological space. Analytic
spaces are topological spaces in their own right with a
countable and point separating base for their topology. As
topological spaces they carry
the $\sigma$-algebra of Borel sets. For the rest of the paper we will
assume that analytic spaces are equipped with just these Borel
sets. This will render notation lighter as well, because it will
permit us to omit the $\sigma$-algebra for an analytic space from notation. Measurability refers to the
Borel sets, unless otherwise noted.

Analytic spaces have a number of
desirable technical properties~\cite{Srivastava, EED-Companion}, among them the closure under
countable products; we note that $\Borel{F\times H} =
  \Borel{F}\otimes\Borel{H}$ for analytic spaces $F$ and
  $H$, $\Borel{\dots}$ denoting the Borel sets. Alas, that the product of
  Borel sets equals the Borel sets of a product is far from being
  common among topological spaces. In general this requires some
  additional assumptions. Just to emphasize this property, we have for
  the analytic spaces $F$ and $H$
  \begin{align*}
    \Borel{F\times H}  &\stackrel{(*)}{=} \sigma\bigl(\{W \mid W\subseteq F \times H \text{
                         is open}\}\bigr)\\    
      &\stackrel{(+)}{=}\sigma\bigl(\{U \times V \mid U\in\Borel{F},
        V\in\Borel{H}\}\bigr)\\
        & =\Borel{F}\otimes\Borel{H}
  \end{align*}
Here equation $(*)$ derives from the definiton of the Borel sets as the
smallest $\sigma$-algebra containing the open sets, and equation $(+)$
derives from the definition of the product $\sigma$-algebra.
  Analytic spaces are also closed under factoring through small
  equivalence relations (\cite[Exercise 5.1.14]{Srivastava},
  \cite[Proposition 4.4.22]{EED-Companion}).

  A first witness to usefulness is given by the following observation
  (cp.~\cite[Corollary~2.11]{EED-Alg-Prop_effFncts}).

\BeginLemma{prod-of-equiv}
  Assume that $\xi$ and $\zeta$ are small equivalence relations on the
  analytic spaces $F$ esp. $H$. Then
  \begin{enumerate}
  \item\label{prod-of-equiv-eins}
    $[\Borel{F\times H}, \xi\times\zeta] = [\Borel{F},
    \xi]\otimes[\Borel{H}, \zeta]$.
\item\label{prod-of-equiv-zwei}
  The measurable spaces
  $\Faktor{(F\times H)}{(\xi\times\zeta)}$ and
  $\Faktor{F}{\xi}\times\Faktor{H}{\zeta}$ are isomorphic.
\end{enumerate}
\EndLemma

Writing down the second assertion in its full beauty means that
$\Faktor{\bigl(F\times H, \Borel{F\times H}\bigr)}{(\xi\times\zeta)}$ is isomorphic
to $\Faktor{\bigl(F, \Borel{F}\bigr)}{\xi}\otimes \Faktor{\bigl(H,
  \Borel{H}\bigr)}{\zeta}$.

\BeginProof
1.
Assume that $\xi$ and $\zeta$ have the respective generators $(U_{n})_{n\in\Nat}$
and $(V_{m})_{m\in\Nat}$. Since
\begin{equation*}
  \isEquiv{\langle x, y\rangle}{\langle x',
    y'\rangle}{(\xi\times\zeta)}
  \Leftrightarrow
  \forall n\in\Nat\forall m\in\Nat:
  \bigl[x\in U_{n}\Leftrightarrow x'\in U_{n}\bigr]
  \wedge
  \bigl[z\in V_{m}\Leftrightarrow z'\in V_{m}\bigr] 
\end{equation*}
we see that
\begin{multline*}
  [\Borel{F\times H}, \xi\times\zeta]
  =\\
  \sigma(\{U_{n}\times V_{m} \mid n, m\in \Nat\})
  =
  \sigma(\{U_{n} \mid n\in \Nat\})\otimes \sigma(\{V_{m} \mid m\in \Nat\})
  =\\
  [\Borel{F},\xi]\otimes[\Borel{H}, \zeta]
\end{multline*}
2. It is not difficult to see that
$\Klasse{\langle x, y\rangle}{\xi\times\zeta} \mapsto \langle
\Klasse{x}{\xi}, \Klasse{y}{\zeta}\rangle$
is a bijection and measurable. Now look a the inverse $\ell$. We want to show that
$\InvBild{\ell}{E}\in \Faktor{\bigl(F, \Borel{F}\bigr)}{\xi}\otimes \Faktor{\bigl(H,
  \Borel{H}\bigr)}{\zeta}$ for each $E\in \Faktor{\bigl(F\times H,
  \Borel{F\times H}\bigr)}{(\xi\times\zeta)}$. By the observation
 following~(\ref{INV}) it is sufficient to show that
  $
  \InvBild{\ell}{\Bild{\eta_{\xi\times\zeta}}{D}}\in
  \Borel{\Faktor{F}{\xi}}\otimes \Borel{\Faktor{H}{\zeta}}
  $
  for every $D\in[\Borel{F\times H}, \xi\times \zeta]$.
      The set
    \begin{equation*}
      \mathcal{D} := \bigl\{D\in [\Borel{F\times H}, \xi\times \zeta] \mid
      \InvBild{\ell}{\Bild{\eta_{\xi\times\zeta}}{D}}\in
      \Borel{\Faktor{F}{\xi}}\otimes \Borel{\Faktor{H}{\zeta}}\bigr\}
    \end{equation*}
certainly contains all rectangles $P\times Q$ with $P\in[\Borel{F},
\xi]$ and $Q\in[\Borel{H}, \zeta]$ and, because the complement of
an invariant set is invariant again, it is closed under
complementation. Also, $\mathcal{D}$ is closed under disjoint
countable unions. Since the set of rectangles with invariant sides is
closed under intersection, Dynkin's celebrated
$\pi$-$\lambda$-Theorem~\cite[Theorem 1.6.30]{EED-Companion}
together with part~\ref{prod-of-equiv-eins} tells us that
$
{\cal D} = [\Borel{F\times H}, \xi\times \zeta].
$
\EndProof

This result is not only of structural importance, as we will see in a
moment. It will also permit us to use, e.g., $\langle
\Klasse{x}{\xi}, \Klasse{y}{\zeta}\rangle$  and  $\Klasse{\langle x,
  y\rangle}{\xi\times\zeta}$ interchangeably, similarly with maps. This will simplify notation
somewhat and thus make life a bit easier.

From now on all automata are working over analytic spaces.

A decent morphism generates a congruence via its
kernel~\cite{Burris-UnivAlgebra, Graetzer}. The following
counterpart to Proposition~\ref{factor-is-morph} shows that this is also
the case with stochastic automata.

\BeginProposition{congr-with-morph}
Given the stochastic automata $\mathbf{K}$  and $\mathbf{K}'$ with $(f, g, h): \mathbf{K}\to  \mathbf{K}'$ an automata
morphism. Then $(\Kern{f}, \Kern{g}, \Kern{h})$ is a congruence for
$\mathbf{K}$, provided $g$ and $h$ are final. .
\EndProposition

\BeginProof
1. Write $\mathbf{K} = \bigl(X, Y, Z, K\bigr)$  and $\mathbf{K}' = \bigl(X', Y',
Z', K'\bigr)$.  We show first that we can find for $V\in[\Borel{Z},
\Kern{h}]$ a Borel set $V_{0}\in\Borel{Z'}$ such that $V =
\InvBild{h}{V_{0}}$, and for $W\in[\Borel{Y}, \Kern{g}]$ another Borel
set $W_{0}\in\Borel{Y'}$ with $W=\InvBild{g}{W_{0}}$. This is done
exactly as in Example~\ref{morpf} using finality of the respective
maps.

2. Assume $f(x) = f(x')$ and $h(z) = h(z')$, and take
$G\in[\Borel{Z\times Y}, \Kern{h}\times\Kern{g}]$. We want to show
that $K(x, z)(G) = K(x', z')(G)$ holds. Assume first that
    $G=V\times W$ with $V\in[\Borel{Z}, \Kern{h}]$ and
    $W\in[\Borel{Y}, \Kern{g}]$
    and determine $V_{0}, W_{0}$ as above, so that $G=\InvBild{(h\times
      g)}{V_{0}\times W_{0}}$. But now
\begin{equation*}
K(x, z)(G) = K(x, z)\bigl(\InvBild{(h\times g)}{V_{0}\times
  W_{0}}\bigr)
= K'(f(x), h(z))\bigl(V_{0}\times W_{0}\bigr)
= K(x', z')(G).
\end{equation*}
This argument shows that
\begin{equation*}
  {\mathcal D} := \bigl\{G\in[\Borel{Z\times Y}, \Kern{h}\times\Kern{g}] \mid
  K(x, z)(G) = K(x', z')(G)\bigr\}
\end{equation*}
contains all rectangles $V\times W$ with $V\in [\Borel{Z}, \Kern{h}]$
and $W\in[\Borel{Y}, \Kern{g}]$. The set of these rectangles is closed
under finite intersections, and ${\mathcal D}$ is closed under complementation as
well as under countable disjoint unions. By Dynkin's
$\pi$-$\lambda$-Theorem ${\cal D}$ equals 
$[\Borel{Z}, \Kern{h}]\otimes[\Borel{Y}, \Kern{g}]$, which is equal to
$[\Borel{Z\times Y}, \Kern{h}\times\Kern{g}]$ by the first part of
Lemma~\ref{prod-of-equiv}.

3. We have shown that $\Kern{f}\times\Kern{h}$ is a subset of
$\Kern{\SubProb{m_{\Kern{h}\times\Kern{g}}}\circ K}$, which
establishes the claim by Lemma~\ref{charact}.
\EndProof

Recall that a map $f: F\to H$ has an em-factorization $f =
\Dt{f}\circ \eta_{\Kern{f}}$. If $f$ is measurable, so are the
components (but this does not entail the em-factorization living in the
category of measurable spaces). We obtain a similar
decomposition for stochastic automata: Let $\mathfrak{f} = (f, g, h):
\mathbf{K}\to \mathbf{K}'$ be a morphism. To express this in a concise
manner, put $\eta_{\Kern{\mathfrak{f}}} :=
(\eta_{\Kern{f}}, \eta_{\Kern{g}}, \eta_{\Kern{h}})$ and $\Dt{\mathfrak{f}} :=(\Dt{f},
\Dt{g}, \Dt{h})$.

We immediately obtain as a consequence of Proposition~\ref{congr-with-morph}:

\BeginCorollary{factor-em}
In the notation of Proposition~\ref{congr-with-morph}, 
$
\eta_{\Kern{\mathfrak{f}}}: \mathbf{K}\to \mathbf{K}_{\Kern{\mathfrak{f}}}
$
and 
$
\Dt{\mathfrak{f}}: \mathbf{K}_{\Kern{\mathfrak{f}}}\to \mathbf{K}'
$
are morphisms, and $\mathfrak{f} = \Dt{\mathfrak{f}}\circ \eta_{\Kern{\mathfrak{f}}}$.
\EndCorollary

\BeginProof
The first part follows from Proposition~\ref{congr-with-morph}
together with Proposition~\ref{factor-is-morph}. As for the second
part, a somewhat lengthy but straightforward computation shows that
\begin{equation*}
  \bigl(\SubProb{\Dt{h}\times\Dt{g}}\circ
  K_{\Kern{\mathfrak{f}}}\bigr)(\Klasse{x}{\Kern{f}},
  \Klasse{z}{\Kern{h}})(E')
 =
 \bigl(K'\circ (\Dt{f}\times\Dt{h})\bigr)(\Klasse{x}{\Kern{f}},
  \Klasse{z}{\Kern{h}}(E')
\end{equation*}
whenever $E'\in\Borel{Z'\times Y'}$. The last equation is obvious.
\EndProof

\section{Sequential Work}
\label{sec:sequential-work}

A stochastic automaton works sequentially and synchronously: input is
fed into it, in each step an output is produced, then a new input is
given, a new output is produced, etc. Of course, state changes occur
as part of these operations. Formally, suppose the automaton $\mathbf{K} = (X, Y,
Z, K)$ is in state $z$ and receives first $x_{1}$, then $x_{2}$ as the
input. Quite apart from the salient state changes, an output of length
two is produced, and the probability for the measurable set
$E\subseteq Z\times Y\times Y$ is computed so:
\begin{equation}
  \label{eq:4}
  K(x_{1}x_{2}, z)(E) := \int_{Z\times Y} K(x_{2}, z')(\{\langle z'', y_{2}\rangle
    \mid \langle z'', y_{1}y_{2}\rangle \in E\})~dK(x_{1}, z)(\langle
  z', y_{1}\rangle) 
\end{equation}
After input $x_{1}$ in state $z$ the automaton makes a transition to
state $z'$ and gives an output $y_{1}$ with probability $dK(x_1, z)(\langle
  z', y_{1}\rangle)$. The new input $x_{2}$ is met in state $z'$
  and produces a new state $z''$ as well as an output $y_{2}$ so that
  $\langle z'', y_{1}y_{2}\rangle\in E$ with probability
  $
   K(x_{2}, z')(\{\langle z'', y_{2}\rangle
   \mid \langle z'', y_{1}y_{2}\rangle \in E\})
   $
We have to average over $z'$ and $y_{1}$. Standard arguments~\cite{EED-LNCS81}
show that
we have extended the transition law to a stochastic relation $K: X^{2}\times Z \Rightarrow Z\times Y^{2}$
(we could use indices showing the length of the automaton's work so
far, but there is already enough notation around).

Let $v\in X^{n}$ be an input word of length $n$, and assume that we
have extended the transition law already to a stochastic relation $K:
X^{n}\times Z\Rightarrow Z\times Y^{n}$, all products carrying the
corresponding product $\sigma$-algebras. Define for the input $x\in
X$, and the state $z$ for the Borel set $E\subseteq Z\times Y^{n+1}$
\begin{equation}
  \label{eq:4a}
  K(vx, z)(E) := \int_{Z\times Y}
  K(x, z')(\{\langle z'', y\rangle \mid \langle z'', wy\rangle\in E\})~dK(v, z)(\langle z', w\rangle).
\end{equation}
Then it is shown in~\cite{EED-LNCS81} that $K: X^{n+1}\times
Z\Rightarrow Z\times Y^{n+1}$ is a stochastic relation.

In this way we extend the probabilistic transition law to finite input
sequences in a natural manner.

Now assume that $\mathfrak{c} = (\alpha, \beta, \gamma)$ is a
congruence for $\mathbf{K}$. We will show now that friendship is not
lost during the automata's sequential work as outlined above.
Define for the equivalence relation $\alpha$ on $X$ and for $n\in
\Nat$ the extension $\alpha^{n}$ of $\alpha$ to $X^{n}$ in the obvious manner
\begin{equation*}
  \isEquiv{\langle x_{1}, \dots, x_{n}\rangle}{\langle x'_{1}, \dots,
    x'_{n}\rangle}{\alpha^{n}}
  \Leftrightarrow \isEquiv{x_{i}}{x'_{i}}{\alpha} \text{ for } i = 1, \dots, n, 
\end{equation*}
similarly for the other equivalence relations, and for, e.g., 
$\alpha^{\infty}$ when dealing with infinite sequences.
We claim that $\alpha^{n}\times\gamma$ is friendly to
$\gamma\times\beta^{n}$ for each $n\in \Nat$, so that congruence
$\mathfrak{c}$ induces an infinite sequence of
friendships. This will be demonstrated for $n=2$ now, the general case
is shown exactly in the same way using induction and
eq.~(\ref{eq:4a}).

We do these steps:
\begin{description}
\item[Step 1:] The set
  $
  \{\langle z, y'\rangle \mid \langle z', y, y'\rangle\in E\}
  $
is a member of $[\Borel{Z}\otimes\Borel{Y}, \gamma\times\beta]$  for each $E\in[\Borel{Z}\otimes\Borel{Y^{2}},
\gamma\times\beta^{2}]$ and for each $y\in Y$. It is easy to see
that the set in question is a Borel set, and because $E$ is
$\gamma\times\beta^{2}$ invariant, and $\beta$ is a reflexive
relation, the set is also $\gamma\times\beta$-invariant.
\item[Step 2:]Let $E\in[\Borel{Z}\otimes\Borel{Y^{2}},
  \gamma\times\beta^{2}]$ and fix $\bar{x}\in X, \bar{y}\in Y$, then
  the map
  \begin{equation*}
    \langle z, y\rangle\mapsto K(\bar{x}, z)\bigl(\{\langle z'',
    \bar{y}\rangle \mid \langle z'', y, \bar{y}\rangle\in E\}\bigr)
  \end{equation*}
  is $[\Borel{Z}\otimes\Borel{Y},
  \gamma\times\beta]$-measurable. Assume that $E$ is a measurable
  rectangle, say, $E = C_{1}\times B_{1}\times B_{2}$, then
  $
  K(\bar{x}, z)\bigl(\{\langle z'',
  \bar{y}\rangle \mid \langle z'', y, \bar{y}\rangle\in E\}\bigr) =
  K(\bar{x}, z)(C_{1}\times B_{2})\cdot I_{B_{1}}(x)
$ 
with
  $I_{B_{1}}$ the indicator function of the set $B_{1}$. This
  constitutes certainly a $[\Borel{Z}\otimes\Borel{Y},
  \gamma\times\beta]$-measurable function by Lemma~\ref{charact}. Applying the principle of
  good sets, Dynkin's $\pi$-$\lambda$-Theorem shows that the set of
  all $E$ for which the claim is true is all of $[\Borel{Z}\otimes\Borel{Y^{2}},
  \gamma\times\beta^{2}]$, because the latter $\sigma$-algebra is
  generated by these rectangles, and because of the first part of Lemma~\ref{prod-of-equiv}.
\end{description}

Now we are poised to show that $\alpha^{2}\times\gamma$ and
$\gamma\times\beta^{2}$ are friends. For this, take $E\in[\Borel{Z}\otimes\Borel{Y^{2}},
  \gamma\times\beta^{2}]$ and assume that 
$
\isEquiv{\langle x_{1}x_{2}, z\rangle}{\langle \bar{x}_{1}\bar{x}_{2},
  \bar{z}\rangle}{\alpha^{2}\times\gamma},
$
then we have according to eq~\ref{eq:4}
\begin{align*}
  K(x_{1}x_{2}, z)(E) &= \int_{Z\times Y} K(x_{2}, z')\bigl(\{\langle z'', y_{2}\rangle
    \mid \langle z'', y_{1}y_{2}\rangle \in E\}\bigr)~dK(x_{1}, z)(\langle
                        z', y_{1}\rangle) \\
  & = \int_{Z\times Y} K(x_{2}, z')\bigl(\{\dots\}\bigr)~dK(\bar{x}_{1}, \bar{z}) (\langle
    z', y_{1}\rangle)\\
  & \text{~~~~(Corollary~\ref{inv-under-integral}, since the integrand
    is $[\Borel{Z}\otimes\Borel{Y},
    \gamma\times\beta]$-measurable)}\\
& = \int_{Z\times Y}  K(x_{2}, z')\bigl(\{\dots\}\bigr)~dK(\bar{x}_{1}, \bar{z}) (\langle
                                         z', y_{1}\rangle)\\
  &\text{~~~~(by Step 1, because $\{\dots\}$ is in  $[\Borel{Z}\otimes\Borel{Y},
    \gamma\times\beta]$)}\\
  & = K(\bar{x}_{1}\bar{x}_{2}, \bar{z})(E)
\end{align*}
Now the claim is established by Lemma~\ref{charact}.

Summarizing, we obtain

\BeginProposition{more-friends}
Let  $(\alpha, \beta, \gamma)$ be a countably generated congruence for the stochastic automaton
$\mathbf{K}$ over analytic spaces. Then $\alpha^{n}\times\gamma$ is friendly to
$\gamma\times\beta^{n}$ for every $n\in\Nat$.\QED
\EndProposition

In what follows, we will deal with finite or infinite sequences of
inputs resp. outputs. Denote as usual for a set $M$ by $M^{+}$ the set
of all finite non-empty words with letters taken from $M$, $|v|$ denotes the
length of word $v\in M^{+}$. $M^{\infty}$ is the set of all infinite
sequences, and $M^{\leq\infty} := M^{+}\cup M^{\infty}$ are all
non-empty finite or infinite sequences over $M$. For $\tau\in
M^{\infty}$ the first $n$ letters are denotes by $\tau_{n}$. If $M$ carries a
$\sigma$-algebra $\mathcal{M}$, $M^{n}$ carries for $n\leq\infty$ the $n$-fold product
$\mathcal{M}^{n}$, and $M^{+}$ the coproduct $\mathcal{M}^{+}$ of
$(\mathcal{M}^{n})_{n\in\Nat}$, finally $M^{\leq\infty}$ has the
coproduct of $\mathcal{M}^{+}$ and $\mathcal{M}^{\infty}$.

Having thus fixed notation, we turn to automata again. Viewed from the outside, a learning system, or a reactive one,
receives an input and responds through an output, the internal states
are hidden from the observer. They are usually assumed to follow some
initial probability distribution $\mu\in\SubProb{Z}$. So we put
\begin{equation*}
  K^{|v|}_{\mu}(v)(G) := \int_{Z}K(v, z)(Z\times G)~d\mu(z)
\end{equation*}
with $v\in X^{+}, G\in \Borel{Y^{|v|}}$, thus $K^{n}_{\mu}(v)$
specifies the probability distribution of outputs of length $n$
given input $v$ with $|v|=n$, provided the initial states are distributed according
to $\mu$. Note that the state changes after each input are
recorded through $K$, but are kept hidden behind a kind of smoke screen
(indicated by computing the probability $K(v, z)(Z\times G)$, hence
not betraying which new state is adopted specifically). Finally, define 
\begin{equation*}
K^{+}_{\mu}(v)(G) := K^{|v|}_{\mu}(v)(G\cap Y^{|v|})
\end{equation*}
(with $G\in\Borel{Y^{+}}$) as the \emph{black box}
associated with the stochastic automaton $\mathbf{K}$.

We note for later use
\BeginLemma{inf-setting}
For every $\mu\in\SubProb{Z}$, 
$K^{+}_{\mu}: X^{+}\Rightarrow Y^{+}$; given a countably generated congruence $(\alpha, \beta, \gamma)$ on
$\mathbf{K}$, $\alpha^{n}$ is a friend to $\beta^{n}$ with respect to
$K^{n}_{\mu}$ for each $n\in\Nat$.
\EndLemma

\BeginProof
It is shown first that $K^{n}_{\mu}: X^{n}\Rightarrow Y^{n}$ is a
stochastic relation for each $\mu\in\SubProb{Z}$~\cite[Example 2.4.8, Exercise
4.14]{EED-Companion}. Since $X^{+}$ is the coproduct of the measurable
spaces $(X^{n})_{n\in\Nat}$, the first assertion follows. For the
second one, fix $G\in[\Borel{Y^{n}}, \beta^{n}]$, and assume
  $\isEquiv{v}{v'}{\alpha^{n}}$. We observe $\isEquiv{\langle v,
    z\rangle}{\langle v', z\rangle}{\alpha^{n}\times\gamma}$ for all
  $z\in Z$, so in particular $K(v, z)(Z\times G) = K(v', z)(Z\times
  G)$, because $\alpha^{n}\times\gamma$ is friendly to
  $\gamma\times\beta^{n}$ by
  Proposition~\ref{more-friends}. Integrating with respect to
  $\mu\in\SubProb{Z}$ yields $K^{n}_{\mu}(v)(G) =
  K^{n}_{\mu}(v')(G)$. So the second assertion follows from Lemma~\ref{charact}.
\EndProof

In fact, we may educate our black box to work on infinite sequences in
such a way
that the finite initial parts are respected. To be specific, we claim that we
find a stochastic relation $K^{\infty}_{\mu}$ between $X^{\infty}$ and
$Y^{\infty}$ such that for the cylinder set
$G = G_{n}\times\prod_{m>n}Y$ with $G_{n}\in\Borel{Y^{n}}$
\begin{equation*}
  K^{\infty}_{\mu}(\tau)(G) = K^{n}_{\mu}(\tau_{n})(G_{n})
\end{equation*}
holds. Consequently,  we intend to find $K^{\infty}_{\mu}: X^{\infty}\Rightarrow
Y^{\infty}$ with $K^{n}_{\mu}\circ \pi_{n}^{X} = \SubProb{\pi_{n}^{Y}}\circ
K^{\infty}_{\mu}$ for all $n$, with $\pi_{n}^{\bullet}: \tau\mapsto \tau_{n}$ as the
projection of an infinite sequence to its first $n$ letters. Thus we
want to close the gap in this diagram
\begin{equation*}
\xymatrix{
  X^{\infty}\ar[d]_{\pi^{X}_{n}}\ar @{-->}[rr]&&\SubProb{Y^{\infty}}\ar[d]^{\SubProb{\pi^{Y}_{n}}}\\
X^{n}\ar[rr]_{K_{\mu}^{n}}&&\SubProb{Y^{n}}  
}
\end{equation*}

Evidently this requires the automaton to be fully probabilistic, i.e.,
that $(K(x, z)(Z\times Y) = 1$ always holds. For measure-theoretic
reasons, we need also a topological assumption.

\BeginProposition{proj-limit}
Let $\mathbf{K} = (X, Y, Z, K)$ be a stochastic automaton such that $X$ and $Y$
are Polish spaces, and $Z$ is an analytic space. Then there exists for
each initial distribution $\mu\in\SubProb{Z}$ with $\mu(Z) = 1$ a
uniquely determined stochastic relation $K^{\infty}_{\mu}: X^{\infty}\Rightarrow
Y^{\infty}$ such that $K^{n}_{\mu}\circ \pi_{n} = \SubProb{\pi_{n}}\circ
K^{\infty}_{\mu}$ for all $n\in\Nat$, provided $K(x, Z)(Z\times Y)  =
1$ for all $x\in X, z\in Z$.  
\EndProposition

\BeginProof
Fix $\mu\in\SubProb{Z}$ with $\mu(Z) = 1$. Define $\pi_{m, n}: y_{1}\dots y_{m}\mapsto y_{1}\dots y_{n}$ as
the projection $Y^{m}\to Y^{n}$ for $m>n$, and put for $\tau\in
X^{\infty}$
\begin{equation*}
  L_{k}(\tau)(G) := K^{k}_{\mu}(\tau_{k})(G),
\end{equation*}
whenever $k\in\Nat$ and $G\in\Borel{Y^{k}}$. Then $L_{k}:
X^{\infty}\Rightarrow Y^{k}$ with $L_{k}(\tau)(Y^{k}) = 1$ for all
$\tau$, and
\begin{equation*}
  L_{n}(\tau) = \SubProb{\pi_{m, n}}\bigl(L_{m}(\tau)\bigr)
\end{equation*}
holds for $m>n$. Thus $\bigl(L_{n}(\tau)\bigr)_{n\in\Nat}$ is a
projective system in the sense of~\cite[Definition~4.9.18]{EED-Companion} for every $\tau\in X^{\infty}$. The assertion
now follows from~\cite[Corollary~4.9.21]{EED-Companion}, a mild
variant of the famous Kolmogorov Consistency Theorem. 
\EndProof

Because of its genesis, the stochastic relation $K_{\mu}^{\infty}$
might be called the \emph{projective limit} associated with automaton
$\mathbf{K}$ and distribution $\mu$.

Our black box works also for infinite sequences of inputs,
answering with a uniquely determined distribution on the set of output
sequences. The price to pay for this is on one hand the full
probabilistic nature of the underlying stochastic relation (given the
requirement, this is only too obvious), and on the other hand the
assumption of working in Polish spaces. This, however, cannot be
relaxed, as~\cite[Example~7.7.3]{Bogachev} shows. 

The distribution on infinite output sequences is consistent with its
initial pieces. Suppose you stop the input sequence at point $n$,
then you obtain the corresponding distribution on the outputs of
length $n$. This observation permits us to decorate trees. Call a subset $\mathcal{T}$ of $X^{\leq\infty}$ a \emph{tree} iff it is
prefix free (so if $p\in\mathcal{T}$ and $q$ is a prefix of $p$, then
$q=p$). We interpret the elements of $\mathcal{T}$ and their prefixes
as paths, and we associate to each path in the tree a
probability: If $\vec{x} = x_{1}\dots x_{n}$ is a path of length $n$, then there
is some $p\in\mathcal{T}$ such that $\vec{x}$ is the prefix of
length $n$ to $p$. So assign to $\vec{x}$ the distribution
$$
T(\vec{x}) := \SubProb{\pi_{n}^{Y}}(K^{|p|}_{\mu})(p)
$$
(with $|p|=\infty$, if $p$ is
  infinitely long). This is well defined: if $\vec{x}$ is the
  prefix of $q\in\mathcal{T}$ as well, we have by construction
  $\SubProb{\pi_{n}^{Y}}(K^{|p|}_{\mu})(p)=
  \SubProb{\pi_{n}^{Y}}(K^{|q|}_{\mu})(q)$. In fact, being the prefix
  of more than one path may occur in
  case the tree branches out at some node later on.

Thus $T(v)(G)$ is the probability that the output is a member of
$G\in\Borel{G^{|v|}}$ after input of the finite sequence $v$ into the
tree. If the finite path associated with $v$ ends in the leaf $x$ (so
that $v = wx$ for some $w$, and $v$ is not a prefix of another word in
$\mathcal{T}$), then the probability that the final
output is a member of $G_{0}\in\Borel{Y}$ is just
$T(v)\bigl(Y^{|v|-1}\times G_{0}\bigr)$.

Friendship is maintained also for infinite sequences. We first show
that the extension $\xi^{\infty}$ of a small equivalence relation
$\xi$ on the measurable space $(F, \mathcal{F})$ is small
again. Assume that $\xi$ is created by the countable set 
$\mathcal{U} := \{U_{n} \mid n\in\Nat\}\subseteq\mathcal{F}$, which
we may assume to be closed under finite intersections (otherwise take
$\{\bigcap_{i\in S}U_{i} \mid \emptyset\not=S\subseteq\Nat \text{ finite}\}$ as a countable creator). Put
\begin{equation*}
  \mathcal{D}_{\mathcal{U}, \xi} := \{\prod_{i=1}^{k}U_{n_{i}}\times\prod_{m>k}F \mid
  n_{i}\in\Nat\text{ for } 1 \leq i\leq k, k\in\Nat\},
\end{equation*}
then it is easy to see that this countable set creates
$\xi^{\infty}$. Note that $\mathcal{D}_{\mathcal{U}, \xi}$ is also closed
under finite intersections. Now let $\mathcal{V}$ be a countable creator for
$\beta$. Assume that $\isEquiv{\tau}{\tau'}{\alpha^{\infty}}$, fix
$\mu\in\SubProb{Z}$ with $\mu(Z)=1$ as before, and let
$G\in\mathcal{D}_{\mathcal{V}, \beta}$, then $G =
prod_{i=1}^{k}V_{i}\times\prod_{m>k}Y$ for some $V_{1}, \dots,
V_{k}\in\mathcal{V}$. Hence $G_{0} :=
\prod_{i=1}^{k}V_{i}\in[\Borel{Y^{k}}, \beta^{k}]$, and
  $\isEquiv{\tau_{k}}{\tau_{k}'}{\alpha^{k}}$, so that we have 
  \begin{equation*}
    K_{\mu}^{\infty}(\tau)(G) = K_{\mu}^{k}(\tau_{k})(G_{0})
\stackrel{(\ddagger)}{=} K_{\mu}^{k}(\tau_{k}')(G_{0}) = K_{\mu}^{\infty}(\tau')(G).
  \end{equation*}
Equality $(\ddagger)$ is implied by the friendship of $\alpha^{k}$ to
$\beta^{k}$ (Lemma~\ref{inf-setting}). Thus $K_{\mu}(\tau)$ agrees
with $K_{\mu}(\tau')$ on $\mathcal{D}_{\mathcal{V}, \beta}$, so these
measures agree on $[\Borel{Y^{\infty}}, \beta^{\infty}] = \sigma(\mathcal{D}_{\mathcal{V}, \beta})$ by Dynkin's
$\pi$-$\lambda$-Theorem (see~\cite[Lemma~1.6.31]{EED-Companion}).

We have shown
\BeginProposition{all-friends}
Let $\mathbf{K} = (X, Y, Z, K)$ be a stochastic automaton with $K(x, Z)(Z\times Y)  =
1$ for all $x\in X, z\in Z$. Assume that $X$ and $Y$
are Polish spaces, and $Z$ is an analytic space and initial
distribution $\mu\in\SubProb{Z}$ with $\mu(Z) = 1$. If $(\alpha,
\beta, \gamma)$ is a countably generated congruence on $\mathbf{K}$,
then  $\alpha^{n}$ is friendly to $\beta^{n}$ with respect to
$K^{n}_{\mu}$ for every
$n\in\Nat\cup\{\infty\}$.\QED
\EndProposition

So friendship turns out to be a surprisingly stable relationship,
maintained even through finite and infinite streams. 

\section{Play it again, Sam}
\label{sec:play-it-again}

Assume that we have factored an automaton, and subsequently we need to factor
again the factored automaton. It will turn out that the resulting
automaton may be obtained by factoring once the automaton from which
we started, albeit with a modified congruence. This result will also
enable us to do the reduction iteratively along the multiple
components. 

Given an equivalence relation $\xi$ on a set $F$, and an equivalence
relation $\zeta$ on the set $\Faktor{F}{\xi}$, define
\begin{equation*}
\isEquiv{x}{x'}{(\comb{\xi}{\zeta})} \text{ iff } \isEquiv{\Klasse{x}{\xi}}{\Klasse{x'}{\xi}}{\zeta}, 
\end{equation*}
hence $x$ is related to $x'$ through the new relation $\comb{\xi}{\zeta}$ iff the class
$\Klasse{x}{\xi}$ of $x$ is related to the class $\Klasse{x'}{\xi}$ through
relation $\zeta$. We may think of $\comb{\xi}{\zeta}$ as a lifting
operation (visually, a $\zeta$-class may be
seen as a sea in which $\xi$-classes swim; the $\GlueOpp$ operator
lifts these classes to the level of the base space). It is clear that $\comb{\xi}{\zeta}$ is countably
generated if both $\xi$ and $\zeta$ are.

Define the bijections

\parbox{.5\linewidth}{
  \begin{equation*}
    \varphi_{\xi, \zeta}:
    \begin{cases}
      \Faktor{F}{(\comb{\xi}{\zeta})} & \to
      \Faktor{(\Faktor{F}{\xi})}{\zeta}\\
      \Klasse{x}{\comb{\xi}{\zeta}} & \mapsto
      \Klasse{\Klasse{x}{\xi}}{\zeta}
    \end{cases}
  \end{equation*}
  }
  and
  \parbox{.5\linewidth}{
  \begin{equation*}
    \psi_{\xi, \zeta}:
    \begin{cases}
      \Faktor{(\Faktor{F}{\xi})}{\zeta} & \to
      \Faktor{F}{(\comb{\xi}{\zeta})}\\
      \Klasse{\Klasse{x}{\xi}}{\zeta}&\mapsto
      \Klasse{x}{\comb{\xi}{\zeta}}
    \end{cases}
  \end{equation*}
}

  We obtain this diagram
  \begin{equation*}
\xymatrix{
  F\ar[rr]^{\eta_{\xi}}\ar[d]_{\eta_{\comb{\xi}{\zeta}}}
  &&\Faktor{F}{\xi}\ar[d]^{\eta_{\zeta}}\\
  \Faktor{F}{(\comb{\xi}{\zeta})}\ar@<1ex>[rr]^{\varphi_{\xi, \zeta}}
  &&\Faktor{(\Faktor{F}{\xi})}{\zeta}\ar@<1ex>[ll]^{\psi_{\xi, \zeta}}
}
\end{equation*}
with
\begin{equation}
  \label{eq:1}
  \varphi_{\xi, \zeta}\circ \eta_{(\comb{\xi}{\zeta})} = \eta_{\zeta}\circ \eta_{\xi}
  \text{ and }
  \eta_{\comb{\xi}{\zeta}} = \psi_{\xi, \zeta}\circ \eta_{\zeta}\circ \eta_{\xi}
\end{equation}

Assume our stage is a measurable space, then we obtain

\BeginLemma{bi-meas}
Let $(F, {\cal F})$ be a measurable space, then
$
\Faktor{(F, {\cal F})}{(\comb{\xi}{\zeta})}
$
and
$
\Faktor{\bigl(\Faktor{(F, {\cal F})}{\xi}\bigr)}{\zeta}
$
are isomorphic as measurable spaces. Moreover
$
\Bild{\eta_{\xi}}{G}\in[\Faktor{\cal F}{\xi}, \zeta],
$
provided $G\in[{\cal F}, \comb{\xi}{\zeta}]$. 
\EndLemma

\BeginProof
1.
We show that the bijections $\varphi_{\xi, \zeta} $ and $\psi_{\xi,
  \zeta}$ from above are measurable. From
$
\varphi_{\xi, \zeta}\circ \eta_{\comb{\xi}{\zeta}} =
\eta_{\zeta}\circ \eta_{\xi}
$
we see that
$
\varphi_{\xi, \zeta}\circ \eta_{\comb{\xi}{\zeta}}
$
is measurable, and since $\eta_{\comb{\xi}{\zeta}}$ is final, we conclude measurability
of $\varphi_{\xi, \zeta}$. Similarly, since the composition of final
morphisms is final again, measurability of $\psi_{\xi, \zeta}\circ
\eta_{\zeta}\circ \eta_{\xi}$ by (\ref{eq:1}) implies measurability
of $\psi_{\xi, \zeta}$.

2.
For establishing the second part, we have to show that $\Bild{\eta_{\xi}}{G}$ is $\zeta$-invariant, since clearly
$\Bild{\eta_{\xi}}{G}\in\Faktor{\cal F}{\xi}$ on account of $G\in{\cal F}$ being $\xi$-invariant. Given $t\in\Bild{\eta_{\xi}}{G}$ and $t'$
with $\isEquiv{t}{t'}{\zeta}$, we find $x\in G'$ and $x'$ with $t=\Klasse{x}{\xi}$ and
$t'=\Klasse{x'}{\xi}$. Hence $\isEquiv{t}{t'}{\zeta}$ translates to
$
\Klasse{\Klasse{x}{\xi}}{\zeta} = \Klasse{\Klasse{x'}{\xi}}{\zeta},
$
equivalently
$
\Klasse{x}{\comb{\xi}{\zeta}}=\Klasse{x'}{\comb{\xi}{\zeta}}.
$
Since $x\in G$, and because $G$ is $\comb{\xi}{\zeta}$-invariant, we
find that $x'\in G$ holds, which means $t'\in\Bild{\eta_{\xi}}{G}$, so that the latter
set is $\zeta$-invariant.  
\EndProof

Thus we may and do identify $\xi$ with $\comb{\eins{F}}{\xi}$, and with
$\comb{\xi}{\eins{\Faktor{F}{\xi}}}$.

Given a congruence on a factor automaton, each equivalence relation on
a factored component space generates individually a new equivalence on
the component proper through lifting, as we have seen above. These new
equivalences are countably generated, if their components are. Combining
all these lifted equivalences will yield a congruence, as will be shown now.

In a slight abuse of terminology, call a congruence \emph{countably
generated} (abbreviated \emph{cg}) iff all its components are.
 
\BeginProposition{again-sam}
Let $\mathfrak{c} = (\alpha, \beta, \gamma)$ be a cg congruence on the
stochastic automaton $\mathbf{K}$, and
$\mathfrak{c'} = (\alpha', \beta', \gamma')$ a cg congruence on the
factor automaton $\mathbf{K}_{\mathfrak{c}}$. Then $\comb{\mathfrak{c}}{\mathfrak{c'}} :=
(\comb{\alpha}{\alpha'}, \comb{\beta}{\beta'},
\comb{\gamma}{\gamma'})$ is a cg congruence on $\mathbf{K}$.
\EndProposition

\BeginProof
1. Write $\mathbf{K} = \bigl(X, Y, Z, K\bigr)$. We know already that $\comb{\mathfrak{c}}{\mathfrak{c'}}$ is
countably generated, so we have to show that
$\comb{\alpha}{\alpha'}\times\comb{\gamma}{\gamma'}$ is friendly to
$\comb{\gamma}{\gamma'}\times\comb{\beta}{\beta'}$ ($\GlueOpp$ binds
stronger than $\times$). This is done through Lemma~\ref{charact}.

2.
Let
$
\isEquiv{\langle{x, z}\rangle}{\langle{x',
    z'}\rangle}{(\comb{\alpha}{\alpha'}\times\comb{\gamma}{\gamma'})},
$
we want to show
$  
K(x, z)(G) =
  K(x', z')(G)
$
for all $G\in[\Borel{Z\times Y}, \comb{\gamma}{\gamma'}\times\comb{\beta}{\beta'}]$.
Fix such a set $G$. Because $\alpha'\times\gamma'$ is friendly to
$\gamma'\times\beta'$, we know that
\begin{equation}
\label{ding}
K_{\mathfrak{c}}\bigl(\Klasse{x}{\alpha}, \Klasse{z}{\gamma}\bigr)(H)
=
K_{\mathfrak{c}}\bigl(\Klasse{x'}{\alpha},
\Klasse{z'}{\gamma}\bigr)(H)
\end{equation}
for all $H\in[\Borel{\Faktor{Z}{\gamma}}\otimes\Borel{\Faktor{Z}{\beta}},
\gamma'\times\beta'].$
From the second part of Lemma~\ref{bi-meas} we see that
$
\Bild{\fMap{\gamma\times\beta}}{G}\in[\Borel{\Faktor{Z}{\gamma}}\otimes\Borel{\Faktor{Y}{\beta}},
\gamma'\times\beta']$. Thus
\begin{align*}
  K(x, z)(G) & = K(x,
               z)\bigl(\InvBild{\fMap{\gamma\times\beta}}{\Bild{\fMap{\gamma\times\beta}}{G}}\bigr)\\
  & = K_{\mathfrak{c}}\bigl(\Klasse{x}{\alpha},
    \Klasse{z}{\gamma}\bigr)\bigl(\Bild{\fMap{\gamma\times\beta}}{G}\bigr)\\
  & \stackrel{(\ref{ding})}{=} K_{\mathfrak{c}}\bigl(\Klasse{x'}{\alpha},
    \Klasse{z'}{\gamma}\bigr)\bigl(\Bild{\fMap{\gamma\times\beta}}{G}\bigr)\\
  & =  K(x', z')(G),
\end{align*}
and we are done.
\EndProof

Factoring twice, each time with a countably generated congruence, has
---~up to isomorphism~--- the same effect as factoring once through a suitably
constructed congruence. This observation is similar to the Third
Isomorphism Theorem in Group Theory~\cite[Corollary~5.10]{Hungerford3}, which tells us what happens when
factoring iteratively through normal subgroups.

\BeginProposition{isomorph-thm}
Let $\mathbf{K}$ be a stochastic automaton with a
countably generated congruence $\mathfrak{c}$, and $\mathfrak{c'}$ a
countably generated congruence on the factor automaton
$\mathbf{K}_{\mathfrak{c}}$. The factor automaton
 of $\mathbf{K}$ for the congruence $\comb{\mathfrak{c}}{\mathfrak{c'}}$, and
the factor
automaton of $\mathbf{K}_{\mathfrak{c}}$ for the congruence $\mathfrak{c'}$ are isomorphic.
\EndProposition

This could more suggestively and more comprehensively be written as
$
\Faktor{\bigl(\Faktor{\textbf{K}}{\mathfrak{c}}\bigr)}{\mathfrak{c'}}
=
\Faktor{\textbf{K}}{(\comb{\mathfrak{c}}{\mathfrak{c'}})}.
$

\BeginProof
0.
We assume that the automaton $\mathbf{K}$ is defined over the analytic spaces $X$,
$Y$, and $Z$, and that the congruences are $\mathfrak{c} =
(\alpha,\beta, \gamma)$ resp. $\mathfrak{c'} = (\alpha',
\beta', \gamma')$. Denote by $\mathbf{K}_{1}$ the factor automaton
 of $\mathbf{K}$ for the congruence $\comb{\mathfrak{c}}{\mathfrak{c'}}$, and
 by $\mathbf{K}_{2}$ the factor
automaton of $\mathbf{K}_{\mathfrak{c}}$ for the congruence
$\mathfrak{c'}$. $K$ is assumed to be the transition law for automaton
$\mathbf{K}$, $K_{i}$ the one for $\mathbf{K}_{i}$, $i = 1, 2$. 

1.
The candidates for the isomorphism are the suspects already indicated in
the equations~(\ref{eq:1}), specifically

\parbox{.5\linewidth}{
\begin{align*}
  \hin{a}: \Klasse{x}{\comb{\alpha}{\alpha'}} & \mapsto
                                                \Klasse{\Klasse{x}{\alpha}}{\alpha'},\\
  \hin{b}: \Klasse{y}{\comb{\beta}{\beta'}} & \mapsto
                                              \Klasse{\Klasse{y}{\beta}}{\beta'},\\
  \hin{c}: \Klasse{z}{\comb{\gamma}{\gamma'}} & \mapsto
                                                \Klasse{\Klasse{x}{\gamma}}{\gamma'},\\  
\end{align*}
}
and
\parbox{.5\linewidth}{
\begin{align*}
  \her{a}:   \Klasse{\Klasse{x}{\alpha}}{\alpha'} & \mapsto \Klasse{x}{\comb{\alpha}{\alpha'}}, 
                                              \\
  \her{b}: \Klasse{\Klasse{y}{\beta}}{\beta'} & \mapsto
                                             \Klasse{y}{\comb{\beta}{\beta'}},\\
  \her{c}: \Klasse{\Klasse{x}{\gamma}}{\gamma'} & \mapsto
                                               \Klasse{z}{\comb{\gamma}{\gamma'}}.\\  
\end{align*}
}

2.
We show that this diagram commutes
\begin{equation*}
\xymatrix{
  \Faktor{X}{(\comb{\alpha}{\alpha'})}\times\Faktor{Z}{(\comb{\gamma}{\gamma'})}
  \ar[rr]^{K_{1}}\ar[d]_{\hin{a}\times\hin{c}}
  &&
  \SubProb{\Faktor{Z}{(\comb{\gamma}{\gamma'})}\times\Faktor{y}{(\comb{\beta}{\beta'}})}
  \ar[d]^{\SubProb{\hin{c}\times\hin{b}}}\\
\Faktor{(\Faktor{X}{\alpha})}{\alpha'}\times
\Faktor{(\Faktor{Z}{\gamma})}{\gamma'}
\ar[rr]_{K_{2}}
&&
\SubProb{(\Faktor{\Faktor{Z}{\gamma})}{\gamma'}\times \Faktor{(\Faktor{y}{\beta})}{\beta'}}
}
\end{equation*}
For this, fix $J\in\Borel{(\Faktor{\Faktor{Z}{\gamma})}{\gamma'}\times
  \Faktor{(\Faktor{y}{\beta})}{\beta'}}$. An easy manipulation shows
that
\begin{equation}
  \label{eq:2}
  \InvBild{(\fMap{\comb{\gamma}{\gamma'}}\times\fMap{\comb{\beta}{\beta'}})}{\InvBild{(\hin{c}\times\hin{b})}{J}}
  =
  \InvBild{(\fMap{\gamma}\times\fMap{\beta})}{\InvBild{(\fMap{\gamma'}\times\fMap{\beta'})}{J}}.
\end{equation}
But now we obtain
\begin{align*}
  K(\hin{a}(\Klasse{x}{\comb{\alpha}{\alpha'}}),
  \hin{c}(\Klasse{z}{\comb{\gamma}{\gamma'}})(J)
  & =
    K_{2}(x,
    z)\bigl(\InvBild{(\fMap{\gamma}\times\fMap{\beta})}{\InvBild{(\fMap{\gamma'}\times\fMap{\beta'})}{J}}\bigr)\\
  & \stackrel{~(\ref{eq:2})}{=}
    K(x,
    z)\bigl(\InvBild{(\fMap{\comb{\gamma}{\gamma'}}\times\fMap{\comb{\beta}{\beta'}})}{\InvBild{(\hin{c}\times\hin{b})}{J}}\bigr)\\
  & = K_{1}(\Klasse{x}{\comb{\alpha}{\alpha'}}, \Klasse{z}{\comb{\gamma}{\gamma'}})\bigl(\InvBild{(\hin{c}\times\hin{b})}{J}\bigr).
\end{align*}
Thus the diagram in question commutes indeed. A similar diagram  for
$(\her{a}, \her{b}, \her{c})$ is shown to
be commutative in exactly the same manner. Because all
contributing maps are bijective and measurable by the remarks at the
beginning of this section, we have found the
desired isomorphisms.
\EndProof

This result indicates that a stepwise reduction is possible. Suppose
that we want to first reduce states according to $\gamma$, and then
reduce inputs and outputs through $\alpha$ resp. $\beta$. We observe
that up to isomorphism
\begin{equation*}
  (\alpha, \beta, \gamma) = \comb{(\eins{X}, \eins{Y},
    \gamma)}{(\alpha, \beta, \eins{\Faktor{Z}{\gamma}})}.
\end{equation*}
Reducing inputs and outputs first and then dealing with states gives
rise to a similar isomorphism:
\begin{equation*}
  (\alpha, \beta, \gamma) = \comb{(\alpha, \beta,
    \eins{Z})}{(\eins{\Faktor{X}{\alpha}}, \eins{\Faktor{Y}{\beta}}, \gamma)}.
\end{equation*}

\section{Conclusion and Discussion}
\label{sec:conclusion}

The notion of a congruence for stochastic automata is defined and
investigated, the interplay of congruences with the kernels of
morphisms is briefly shed light on. The central notion is the
friendship of equivalence relations with respect to  stochastic
relations, which is studied extensively. We investigate also the behavior of an
automaton when the input comes from a finite or infinite stream; this
permits to have the automaton work on trees with possibly
infinite paths. Some topological assumptions had
to me made in order to face measure theoretic problems
adequately. Finally an isomorphism result is stated which permits the
reduction of an automaton in a stepwise fashion.

An extension to these ideas extends equivalence relations to act on
subprobabilities. To be specific, let $(F, {\cal F})$ be a measurable
space, $\xi$ an equivalence relation on $F$ with the  $\sigma$-algebra
$[{\cal F}, \xi]$ of $\xi$-equivalent sets. Define for $\mu,
\nu\in\SubProb{F, {\cal F}}$ 
\begin{equation*}
\isEquiv{\mu}{\nu}{\rnd\xi} \text{ iff } \forall E\in[{\cal F}, \xi]: \mu(E) = \nu(E).
\end{equation*}
This is the \emph{randomization} of $\xi$~\cite{EED-CoalgLogic-Book};
note that $\isEquiv{x}{x'}{\xi}$ iff $\isEquiv{\delta_{x}}{\delta_{x'}}{\rnd\xi}$ with $\delta_{x}$ the
point mass on $x$. Furthermore, extend the stochastic
relation $K: (F, {\cal F})\Rightarrow(H, {\cal H})$ to a measurable
map $K^{*}: \SubProb{F, {\cal F}}\to \SubProb{H, {\cal H}}$ upon
setting
\begin{equation*}
 K^{*}(\mu)(E) := \int_{F}K(x)(E)\mu(dx)
\end{equation*}
for $E\in{\cal H}$ (remember, stochastic relation $K$ is really a
Kleisli morphism for the Giry monad, $K^{*}$ is its Kleisli
extension). Call then the equivalence relation  $\xi$ a \emph{random
  friend}\footnote{The present author does not know whether a random
  friend is a casual acquaintance, or a friend for life, or something
  in between.}
to the equivalence relation $\zeta$  iff we have
$
\isEquiv{K^{*}(\mu)}{K^{*}(\nu)}{\rnd\zeta}
$
provided
$
\isEquiv{\mu}{\nu}{\rnd\xi}
$
with $\mu, \nu\in\SubProb{F, {\cal F}}.$ An equivalent formulation
without explicit randomization reads
\begin{equation*}
  \Kern{\SubProb{m_{\xi}}}\subseteq \InvBild{(K^{*}\times K^{*})}{\Kern{\SubProb{m_{\zeta}}}},
\end{equation*}
where $m$ is defined in Lemma~\ref{charact}. Transporting these ideas
to automata, one would have to decide whether one wants friendship of
the level of, say, $\rnd{(\alpha\times\gamma)}$, or of
$\rnd{\alpha}\times\rnd{\gamma}$; the latter one indicates a much
tighter pairing than the former one (recall that a finite measure on a
product space in not necessarily a product measure).

\paragraph{Acknowledgement} The author wants to thank Dr.~Jan Bessai
(TU Dortmund) for his curiosity regarding the behavior of an automaton on trees. 


\vspace{1.0cm}
Dr. Ernst-Erich Doberkat\\
{Walther-von-der-Vogelweide-Str. 46\\
  97422 Schweinfurt\\
  Germany
\end{document}